# Dynamic Response of Ionic Current in Conical Nanopores


Zhe Liu,[1,2] Long Ma,[1] Hongwen Zhang,[1] Jiakun Zhuang,[1] Jia Man,[1] Zuzanna S. Siwy,[4] and Yinghua Qiu[1,2,3*]

1. Key Laboratory of High Efficiency and Clean Mechanical Manufacture of Ministry of Education, National Demonstration Center for Experimental Mechanical Engineering Education, School of Mechanical Engineering, Shandong University, Jinan, 250061, China

2. Shenzhen Research Institute of Shandong University, Shenzhen, 518000, China

3. Suzhou Research Institute of Shandong University, Suzhou, 215123, China

4. Department of Physics and Astronomy, University of California, Irvine, California 92697, United States

*Corresponding author:

yinghua.qiu@sdu.edu.cn,




**Abstract**


Ionic current rectification (ICR) of charged conical nanopores has various applications in fields including nanofluidics, bio-sensing, and energy conversion, whose function is closely related to the dynamic response of nanopores. The occurrence of ICR originates from the ion enrichment and depletion in conical pores, whose formation is found to be affected by the scanning rate of voltages. Here, through time-dependent simulations, we investigate the variation of ion current under electric fields and the dynamic formation of ion enrichment and depletion, which can reflect the response time of conical nanopores. The response time of nanopores when ion enrichment forms i.e. at the "on" state is significantly longer than that with the formation of ion depletion i.e. at the "off" state. Our simulation results reveal the regulation of response time by different nanopore parameters including the surface charge density, pore length, tip, and base radius, as well as the applied conditions such as the voltage and bulk concentration. The response time of nanopores is closely related to the surface charge density, pore length, voltage, and bulk concentration. Our uncovered dynamic response mechanism of the ionic current can guide the design of nanofluidic devices with conical nanopores, including memristors, ionic switches, and rectifiers.


**Keywords:**





# Abstract Graphic

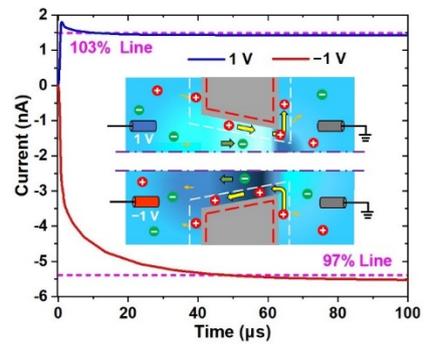



## 1. Introduction

With the development of micro/nanofabrication techniques, nanopores of various sizes and shapes have been successfully manufactured, and widely used in the research and applications of nanofluidics.[1-5] Due to the unique asymmetric morphology, charged conical nanopores exhibit ion current rectification (ICR) behaving like diodes.[6-8] When voltages with opposite polarities are applied across such nanopores, the transport of ions has a preferred moving direction, i.e. the nanopore exhibits the open state under one voltage polarity and the closed state under the other polarity. As the versatile platform for studying the transport of ions and fluid in confined spaces, conical nanopores have potential utilization in ion sieves,[9, 10] ionic circuits,[11-13] seawater desalination,[14, 15] osmotic energy conversion,[16, 17] and bio-sensors.[18, 19]

Based on nanofluidic experiments and simulations, the mechanism and controlling parameters of ICR in nanopores have been well studied. The ICR can be regulated by the nanopore parameters, properties of electrolyte solution, and external experimental conditions.[6, 7, 20-22] The parameters of conical nanopores include the tip and base sizes,[23] cone angle,[24, 25] surface charge density,[26, 27] slip length,[28] hydrophobicity,[29, 30] and surface coating.[31] The properties of solution that influence ICR include salt concentration, pH, viscosity,[32] solvent types,[33] ionic valence,[34] and ionic species.[35] The applied experimental conditions contain the hydrostatic pressure,[36] concentration gradients,[32, 37, 38] and viscosity gradients.[32, 39, 40] In conical nanopores, ICR originates from the ionic selectivity to counterions at the tip region,[41,



[42] which leads to the formation of significant ionic enrichment and depletion inside the nanopores under opposite polarities, corresponding to the obtained larger and smaller ionic currents.[8, 43, 44]

In aqueous solutions, due to electrostatic interactions ions have hydration layers that can move with the ionic transport.[45, 46] Under electric fields, the migration speed of ions has a finite value that can be predicted from the Nernst-Plank equation with the diffusion coefficient.[47] For example, the mobility of $K^+$ ions is $7.96 \times 10^{-8}$ $m^2$/V·s in an infinitely dilute solution under 298 K.[48] When conical nanopores are used as ionic rectifiers, the open and close states are realized by the high and low nanopore conductivity caused by ion enrichment and depletion formed inside nanopores. In earlier reports, the formation of ionic enrichment/depletion regions in conical nanopores was shown to take a considerable amount of time under a transmembrane electric field. With finite element simulations, White et al.[49] found that the ICR in conical nanopores depended on the solution conductivity near the pore tip under different voltage polarities. The maximum and minimum values of solution conductivity occurred at the tip region, ~200 nm from the tip boundary. They reported that it took several milliseconds for the ionic enrichment to form inside nanopores. With nanofluidic experiments, Guerrette et al.[50] found that ICR in glass nanopipettes depended on the scanning rate of applied voltages. With the scanning rate increasing, ICR in conical nanopores became weaker and disappeared completely at high scanning rates such as 200 V/s. The experimental finding was subsequently confirmed by Momotenko et al.[51] using finite element simulations. Under a high



scanning rate of voltage, ions do not have enough time to migrate to form the enrichment and depletion inside nanopores, resulting in linear I-V curves through conical nanopores. With conically shaped nanopores, the memory effects were also observed as loops in the I-V curves recorded under moderate scanning rates of voltage. The memristor properties were attributed to the non-instant (dynamic) redistribution of ions inside nanopores due to the finite mobility of ions.[52-54]

The dynamic response of nanofluidic devices is therefore limited by the migration velocity of ions. Understanding the dynamic response of nanofluidic systems is especially important in applications in ionic circuits and ionic switches.[11, 12, 55] Tybrandt et al.[55] designed a logic ionic circuit composed of npn and pnp ionic transistors to implement the NAND functions. The circuit had a long delay time of seconds due to the slow speed of ionic transport and the size of the system. It was noted that the delay time could be decreased by reducing the transport length of ions or using electrolytes with higher ion mobilities. For the memristors based on conical nanopores of a given geometry, the characteristics were controlled by the salt type, ionic concentration, and solution pH.[56, 57] As important elements in ion circuits, systematic investigation of dynamic characteristics of ion current in conical nanopores under various conditions is crucial to evaluating the dynamic response performance and obtaining regulation rules of conical nanopores as the ionic rectifiers.[50, 51]

Here, stationary and time-dependent models have been constructed by finite element simulation to understand the mechanisms underlying the dynamic responses



of ion current in conical nanopores under various conditions. We studied the influence of the surface charge density, geometrical parameters of conical nanopores, applied voltage, and solution concentration on the required time for the current stabilization in nanopores. Our results show that the dynamic response of ionic current through conical nanopores can be influenced by all the parameters considered. In our electrode configuration, at positive voltages, the base of the conical nanopore has a higher electric potential than the tip causing counterions to move from the base to the tip of nanopores. The time for the current in the pore to reach stabilization is shortened (i.e. the pore has a faster response) by increased surface charge density as well as decreased tip diameter, pore length, and bulk concentration. Interestingly, the response time at negative voltages (i.e. when the tip of the conical nanopore has a higher electric potential than the base) is much longer than that at positive voltages. The negative ion currents can, however, stabilize faster with the decrease of surface charge density, pore length, and voltage, as well as the increase of tip and base diameters, and solution concentration. Our results reveal the physical details of ion migration during the dynamic response of ion current in conical nanopores, which increases the microscopic understanding of ICR and provides theoretical guidance for the design of ion rectifiers, switches, and memristors.

Note that our model describes the dynamic distribution of electrochemical potential related to ionic transport. It cannot predict or describe dynamic processes of nucleation or phase transition, including the formation of nanobubbles, solute nucleation, or precipitate formation.[58-60] These additional processes could lead to more pronounced dynamic effects.[29, 61, 62] While, based on the dependence of the dynamic process on the physical properties of liquids inside nanopores, they can be



evaluated or predicted with the simulations, such as the occurrence of nanoprecipitation due to ion accumulation ion conical nanopores.[58]

## 2. Simulation Details

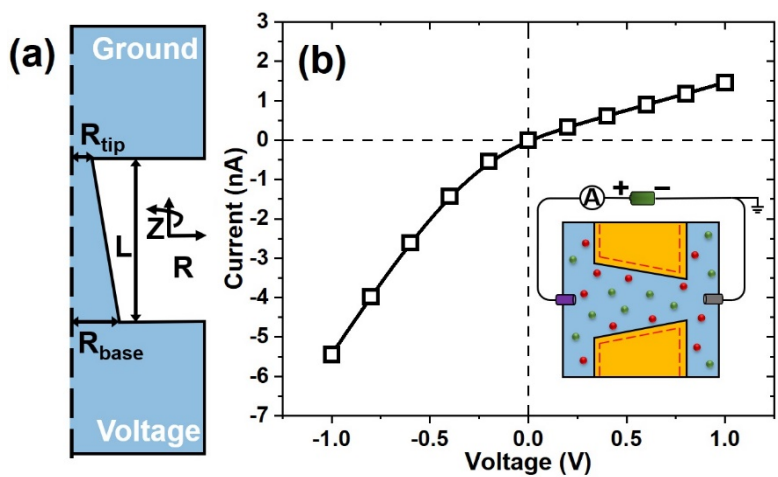

Figure 1. Simulation scheme. (a) Geometric model of the conical nanopore and reservoirs (not to scale). The length as well as tip and base radii of the conical nanopore are denoted as $L$, $R_{tip}$, and $R_{base}$. (b) The current-voltage curve obtained in the conical nanopore. Inset shows the setup of electrodes in the simulation, i.e. the ground and working electrodes are put on the tip and base sides, respectively.

Dynamic and stationary simulations were performed using COMSOL Multiphysics to explore the characteristics of ion current in conical nanopores.[8, 51, 63] Figure 1a shows the schematic diagram of the simulation model which consists of two cylindrical reservoirs connected by a conical nanopore. The radius and length of the reservoir were set to 5 μm. The tip radius ($R_{tip}$) of the conical nanopore was set from 1 to 3 nm, with the default value of $R_{tip}$=2 nm. Additional simulations were also performed with $R_{tip}$=5 nm and $R_{tip}$=7.5 nm. The base radius ($R_{base}$) ranged from 40 to 60 nm, and the default $R_{base}$= 50 nm. The length ($L$) of the conical nanopore varied



from 300 to 700 nm, and the default pore length was 500 nm.[8, 64, 65] Both the inner and outer walls of conical nanopores were negatively charged. The surface charge density ranged from −0.01 to −0.16 C/m$^2$, with a default value of −0.08 C/m$^2$.[63, 66] Considering that KCl had been widely used in nanofluidic experiments,[1, 19] and the anions and cations have similar mobilities, KCl was chosen as the electrolyte. The diffusion coefficients of K$^+$ and Cl$^-$ ions were set as 1.96×10$^{-9}$ and 2.03×10$^{-9}$ m$^2$/s, respectively.[48] Different KCl concentrations from 5 to 200 mM were used, and the default concentration was 100 mM. Voltages across nanopores were applied from −1 V to 1 V. In the system, the permittivity of water was set to 80 and the temperature was constant at 298 K.

In the simulation, the Poisson-Nernst-Planck and Navier-Stokes equations were used to describe the distribution of ions at the solid-liquid interface, and the flow of ions and liquids inside the system. [8, 63, 67] The stationary and time-dependent models were used to study the steady state and dynamic response of ion transport in a conical nanopore under electric fields. Detailed boundary conditions for the simulation are listed in Table S1. Figure S1 shows the construction strategy of mesh.[8, 63, 67] Triangular meshes were selected in the model with a size of 0.1 nm on the inner surface and an area within 2 μm away from the pore boundaries on both outer surfaces. For the remaining area of both outer surfaces, a mesh size of 0.5 nm was applied. In our simulation model, the membrane material was not included to lower the calculation cost. We verified that the results presented in this manuscript are not affected by whether the membrane material is included in the simulations (see



below).

At different voltages, the ionic current through the nanopore was obtained from the integration of ionic flux at the boundary of the reservoir, with Equation 1.[8, 63, 67]

$$I = \int_S F\left(\sum_i^2 z_i \boldsymbol{J}_i\right) \cdot \boldsymbol{n} \, dS \qquad (1)$$

where, $S$, $F$, $z_i$, $\boldsymbol{J}_i$, and $\boldsymbol{n}$ represent the reservoir boundary, Faraday's constant, valence of ionic species $i$ ($K^+$ and $Cl^-$ ions), ionic flux, and unit normal vector, respectively.

## 3. Results and Discussion

Figure 1(b) shows an I-V curve obtained from finite-element simulations for a single conical nanopore in 100 mM KCl. After applying voltages with opposite polarities but equal magnitudes across the nanopore, ICR occurs with a larger ionic current at negative voltages.[7] For the electrode setup as shown in the illustration, at negative voltages, counterions move from the tip to the base of the negatively charged conical nanopore. The larger ionic current in conical nanopores corresponds to the open state of the pore. Many studies have revealed that the high conductance state stems from the high conductivity of a solution inside the nanopore that is caused by the voltage-induced enrichment of ion concentrations. [43, 44] As shown in Figure S2a, at negative and positive voltages the concentration distributions along the pore axis exhibit obvious ionic enrichment and depletion, respectively. At 1 V, the ion concentration is much lower than the bulk concentration over a significant length of the conical nanopore. As the number of ions migrating out of the pore is greater than the number of ions entering the pore, a depletion region is created. At −1 V, on the



other hand, ionic concentrations are enriched with the largest enhancement present at the tip region of the conical nanopore.

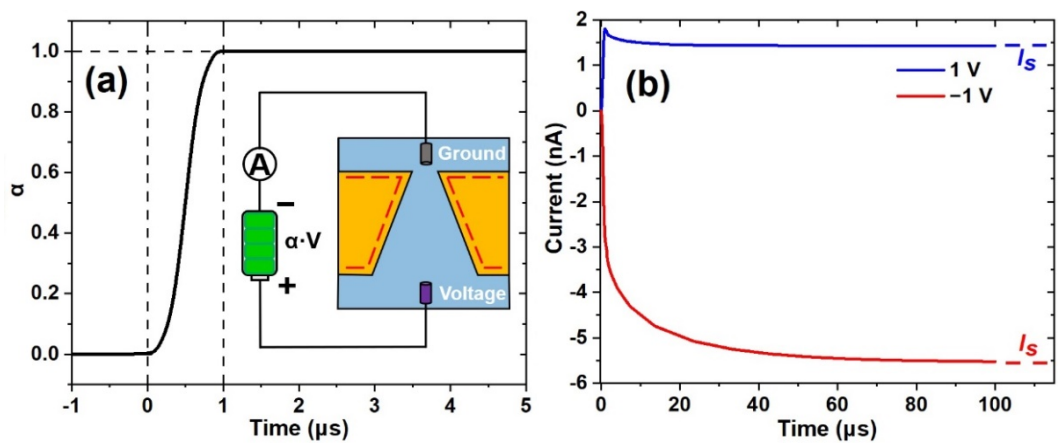

**Figure 2.** Characteristics of dynamic current in the conical nanopore. (a) The step function curve used for the dynamic voltage across the nanopore. The voltage was increased from 0 to the applied value within 1 μs. The simulation before t=0 was conducted to form EDLs on the charged surfaces in the system. (b) Time response curves of ion current in conical nanopore at ±1 V.

Following the work of Guerrette et al.[50] and Momotenko et al.,[51] we conducted both nanofluidic experiments and finite element simulations. From our results, a similar trend in ICR was found under different scanning speeds of voltages (Figure S3).[68] With the increase of the voltage scanning rate, ICR gradually disappeared in the conical nanopore. Subsequently, time-dependent computational models were used to investigate the ion current response in conical nanopores. As shown in Figure 2, step voltages were set in the time-dependent models to simulate the voltage application during the nanofluidic experiment. In COMSOL simulations, the time change of the built-in step function was 1 μs, which can also be set as other values such as 10 μs (Figure S4). The function for the applied voltage across the nanopore



is described by Equation 2.

$$Volt = Volt_0 \cdot \alpha(Time) \qquad (2)$$

where *Volt* is the real-time voltage across the nanopore, $Volt_0$ is the voltage at the final steady-state, and $\alpha$ is the regulation factor which is related to time. The relationship between the step function and time is shown in Figure 2a, where it takes 1 µs for the dependent variable α to increase from 0 to 1. To simulate the real nanofluidic experiments, systems were equilibrated for 1 µs before time 0. During the equilibrium, EDLs were formed on charged surfaces.

Figure 2b shows the dynamic response of the current in a conical nanopore after the application of ±1 V. Within 1 µs, as the applied voltage gradually increases from 0 to the set voltage value, ion current in the nanopore gradually increases as well, to a value that is larger at the negative voltage. When the voltage reaches the set level and remains unchanged, the ion current in the nanopore is still dynamically changing towards the steady-state current values. Under opposite voltage polarities, ionic current exhibits different response times, i.e. different time is needed for the ion current to reach the steady state values at positive and negative voltages. As shown in Figure 2b, the response time of the conical nanopore is significantly longer at −1 V compared to that at +1 V. We would like to note that the asymmetry in the dynamic response is responsible for the memristor properties of conical nanopores, and corresponds to the larger memristor loop at the "on" state of conical nanopores, as reported before.[53, 54, 56, 57] Interestingly, at positive voltages, the current in the nanopore first increases to a peak value and then gradually decreases to the



steady-state value.

We also considered a possible influence of parasitic capacitance on the dynamic response of conical nanopores and the resulting ion transport.[50, 54] Following the strategy of Wang et al.,[53] a series of simulations with membrane radii, modeled as the size of reservoirs, varying from 1 μm to 7 μm were conducted to explore the capacitance effect on the current behaviors through conical nanopores. As shown in Figure S5, under dynamic voltage change we considered, the capacitance has little influence on the response time of conical nanopores. In additional simulations, we considered 5 μm-in-radius PET and glass membranes (Figure S6). The response time of the conical nanopore shares almost the same value as that obtained in the simulation without the membrane material.

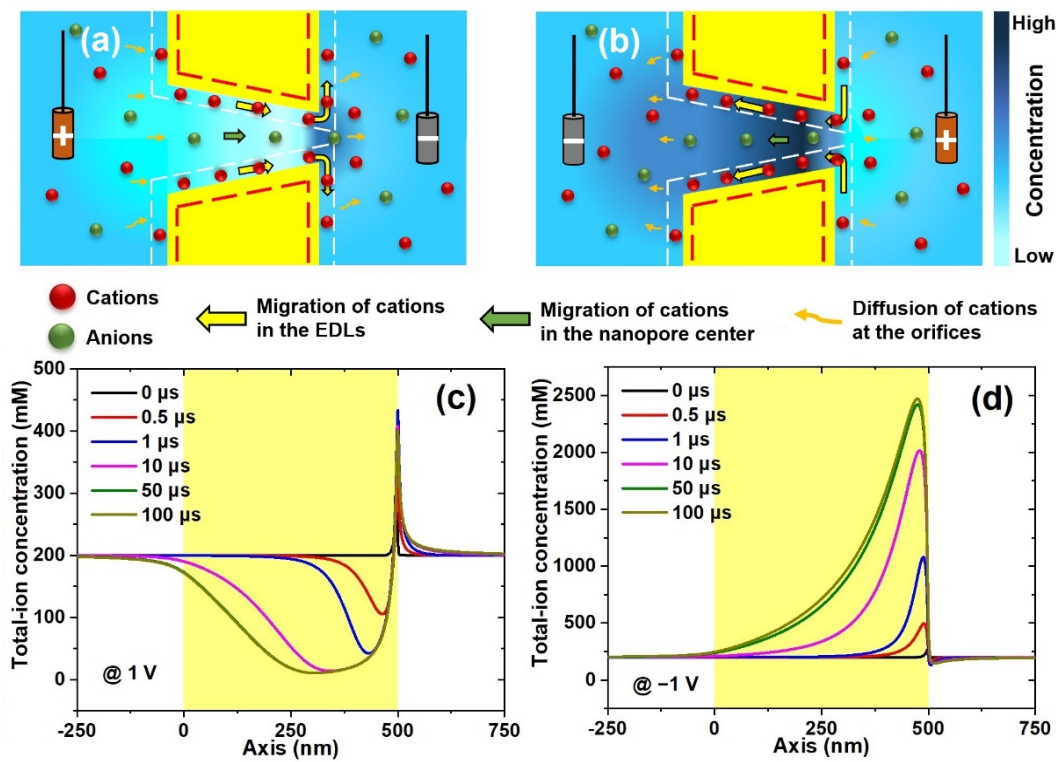

**Figure 3.** Ion distributions in conical nanopores in 100 mM KCl. (a)-(b) Schematic diagram of ion migration under positive voltages (a) and negative voltages (b). (c)-(d)



Variation of total-ion concentration with time along the pore axis of the conical

nanopore at 1 V (c) and −1 V (d). Yellow areas indicate the nanopore region. The

pore length is 500 nm.

Figure 3 shows the ionic migration process inside the conical nanopore under

electric fields. In conical nanopores, the tip region features the largest confinement.[18, 69] The negative charges on pore walls provide strong electrostatic attraction or

repulsion to the cations and anions in the solution, respectively, which induces the ion

selectivity of the nanopore.[8] It is the synergistic action of both the ionic selectivity of

the nanopore and applied electric fields that can lead to ionic enrichment and

depletion inside the charged conical nanopore. Based on our previous research, in an

aqueous solution, EDL regions near the charged walls provide a fast passageway for

counterions.[63, 67, 70] As the main current carriers, $K^+$ ions can enter/exit the negatively

charged conical nanopore in the EDL regions and the center of the pore. However, $Cl^-$

ions migrate mainly through the central region of the conical pore.

When negative voltages are applied, cations enter the conical nanopore from the

tip and exit at the base. At the initial moment of the voltage application, the highly

confined space at the pore tip accounts for the majority of the total resistance in the

system, which leads to a large voltage drop at the tip region.[49] Under the induced

strong electric field, counterions can migrate quickly into the nanopore. Near the base

side of conical nanopores, the decrease in space confinement leads to a reduced

electric field strength. The ionic migration is slowed down which induces the

accumulation of counterions inside the conical nanopore through the effect of



concentration polarization. The high ion concentration resulting from ionic enrichment further reduces the resistance of the local region in the system, resulting in a lower local electric field strength and a smaller speed of ion transport.[71] The positive feedback stabilizes the ion transport, and results in the stable ion enrichment inside the nanopore. Note that outside the pore tip, the rapid ionic transport induces a region of weak ionic depletion where the strong electric field at the pore entrance can promote the migration of counterions along the EDL regions near charged exterior surfaces.[63]

When positive voltages are applied, the high electric field strength inside the pore tip transports counterions quickly outside of the pore. The weak electric field strength at the pore base, on the other hand, weakens the migration of cations into the pore base from the bulk. At the pore base, counterions are mainly driven by ion diffusion and weak ion migration. As the number of ions exiting the pore is at first higher than the number of ions replenishing the pore interior (Figure S8), the ionic depletion regions gradually form inside the nanopore, and reduce the solution conductivity inside the nanopore. This in turn strengthens the local electric field strength and promotes the migration out of the pore of ions in the depletion regions out of the pore until the equilibrium is reached with the formation of a large and stable depletion zone. The enrichment and depletion of ions formed at opposite voltage polarities inside conical nanopores lead to the large and small ion currents, respectively, i.e. the "on" and "off" states of the pore.

Due to the dependence of ionic current on the ion concentration inside



nanopores, the variation of ion concentration with time can shed light on the dynamic processes during current rectification.[41, 51, 63] Figures 3c and 3d show the concentration distributions of total ion concentrations along the pore axis at different times under both voltage polarities. At 0 μs, with no voltage applied, the equilibrium of the system is reached. EDLs form near the charged surfaces of the nanopore,[71] and a clear accumulation of ions at the pore tip is formed. After the voltage is applied across the nanopore, already within 1 μs obvious ion enrichment and depletion occur inside the pore at negative and positive voltages, respectively.[43, 44] After 1 μs, the applied voltages remain constant, while the ion concentration needs more time to reach the steady state.[49] The time required for ion enrichment and depletion to form, from the moment the voltage is applied till the stabilization is reached, affects the dynamic response of ion current in conical nanopores. We also look at how the peak values of ion concentration in the distributions at the pore axis depend on time. Both normalized curves of ion concentration and ionic current are found to share a similar trend which confirms the dependence of ionic current on ion concentration (Figure S8).

   As the next step, we wanted to understand the relationship of our findings to the memristor properties of conical nanopores, as reported before.[53, 54, 56, 57] To this end, we performed similar simulations as shown by Wang et al.[53] and modeled memristor loops in I-V curves under different scanning rates of voltage from 1 V/s to 1 MV/s, as shown in Figure S9. We found that in the "on" state, the longer time needed for the formation of ionic enrichment inside the pore was responsible for the more



pronounced hysteresis in the memristor loop of conical nanopores.[53, 54, 56, 57] The memory effects were observed at the scanning rate of ~10 kV/s, corresponding to a response time of ~100 µs which is very close to our results shown in the manuscript ~50 µs. Consequently, under a specific applied scanning rate of voltage, if the dynamic change of voltage is slower than the response time to form ionic enrichment inside nanopores, the memristor characteristics will not appear in I-V curves.

Taking advantage of current rectification effects, conical nanopores find wide applications in ionic circuits,[11-13] nanofluidic sensors,[18, 19] osmotic energy conversion,[16, 17] and seawater desalination.[14, 15] The realization of the functions is closely related to the ion transport process in nanopores, whose speed determines the dynamic response characteristics of nanopores. Here, we investigated the dynamic response of conical nanopores as a function of various properties of nanopores and external conditions, which not only enriches our understanding of the dynamic transport characteristics of ions but also provides guidance to the design of nanofluidic devices. Here, we assume the stability in time-dependent simulations is reached when the ionic current in time-dependent models becomes 97% of that in stationary models.[8, 63, 70] The time required for this process is denoted as the response time of the nanopore, i.e. the minimum time for the current stabilization (Figures S10 and S11).



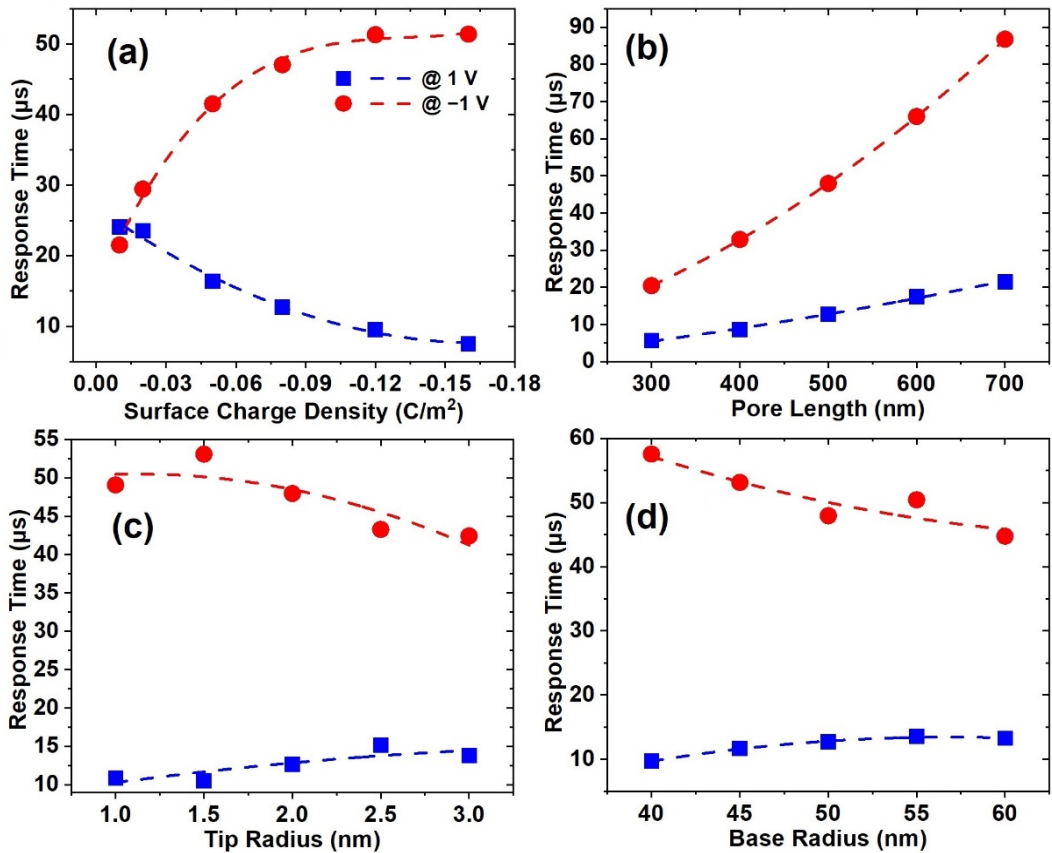

**Figure 4**. Response time of conical nanopores for different geometrical and electrochemical pore parameters. (a) Surface charge density, (b) pore length, (c) tip radius, and (d) base radius.

Figure 4 shows the response time of nanopores, i.e. the time required to reach current stabilization, under the influences of nanopore parameters, such as surface charge density, pore length, tip radius, and base radius. The surface charge density is an important parameter of nanopores, which determines the ion selectivity of the nanopore.[72] The EDL regions near charged pore walls can provide a fast passageway for the transport of counterions. Consequently, surface charge density is expected to have a significant influence on the stabilization time.[63, 70] Figure 4a confirms this hypothesis and shows that at opposite voltage polarities, the response time has very different trends with the variation in the surface charge density. As the



surface charge density increases, nanopores exhibit a stronger selectivity to counterions. In addition, as more counterions need to be accumulated in EDLs to shield surface charges, the current in EDL regions increases under electric fields. At positive voltages, ionic depletion is formed inside the conical nanopore, corresponding to the off state. Due to the enhanced ionic migration in EDL regions[63, 70] caused by the higher surface charge density, a large number of ions migrate rapidly from the tip to the outside of the pore. The current difference collected in the EDL regions[63, 73] between the tip and base shows that the ionic flux exiting the pore is greater than that entering the pore. Figure S12 confirms that this current difference is positively correlated with the surface charge density, and a higher surface charge density can lead to a faster formation of stable ion depletion at positive voltages. Due to the weak dependence of the number of ions at positive voltages on the surface charge density, as the surface charge density increases from $-0.01$ C/m$^2$ to $-0.16$ C/m$^2$, the time for the current to reach the stable state decreases from 24 μs to 7.5 μs. At negative voltages, on the other hand, ion enrichment is formed in the nanopore, corresponding to the "on" state of the system. EDL regions near charged pore walls can transport a large number of counterions into nanopores. Since higher concentrations of ions can form in nanopores with a high surface charge density (Figure S13), the required larger number of ions needs more time to accumulate inside the pore. As the surface charge density increases from $-0.01$ C/m$^2$ to $-0.16$ C/m$^2$, the response time of nanopores increases by ~240%, from 21.5 μs to 51.3 μs. The simulations allowed us to conclude that the longer response time at negative



voltages and higher charge densities can be attributed mainly to the larger number of ions that need to be transported to create the ion enrichment zone, than the number of ions transported for the depletion zone to be formed.

Pore length is another parameter that can affect the dynamic response of conical nanopores because it determines the volume of ion enrichment and depletion inside nanopores. In addition, a longer nanopore has a larger charged surface area, which leads to more significant ion enrichment and depletion. As shown in Figure 4b, the pore length has a clear influence on the response time of ion current. At both positive and negative voltages, the response time of conical nanopores is proportional to the pore length. As shown in Figure S14, as the pore length increases, more ions are required to be enriched or depleted to reach the steady state. Moreover, the decreased electric field strength in long nanopores leads to a reduced migration rate of ions, which results in a slower response of conical nanopores. Since the pore length has a greater impact on the ion enrichment inside nanopores, the response time at the on-state of conical nanopores is more affected by the pore length. As the pore length increases from 300 nm to 700 nm, the response time increases from 20.4 µs to 86.8 µs.

Due to the asymmetric geometry of conical nanopores, both tip and base radius were considered to explore the dependence of pore response time on the pore size. The tip radius determines both the ion selectivity of the nanopore and the area of the EDL regions,[70] which have a trade-off relationship, i.e. when the EDL regions have a large/small area at a large/small tip radius, the ion selectivity of nanopores is



weak/strong. With the tip radius increasing, the degree of ion enrichment and depletion inside the nanopore is weakened (Figures S15a and S15b). In Figure 4c, with the tip radius increasing from 1 to 3 nm, the response time at 1 V changes from ~10 to ~15 μs. In addition, the electric field strength at the pore tip decreases with the increase of the tip radius, which results in reduced ionic migration out of the pore (Figure S15c). When the total number of migrating ions remains constant, a lower migration rate results in a longer time required to reach dynamic equilibrium. At −1 V on the other hand, the response time and the enrichment degree decrease with the increase of the tip radius (Figure S15b). Although the weaker electric field strength leads to a slower ionic migration rate (Figure S15d), the larger area of EDL regions results in considerable ion entry flux, which ensures that the ion enrichment process can reach stability faster.

Note that in highly confined spaces smaller than 3 nm, many other factors like ion size, ionic hydration, and ion-ion interactions may also affect ion distribution and ionic transport.[45, 74, 75] Molecular simulations are usually required to account for these factors. To verify that the results presented here are not affected by ion-specific parameters or ion correlations, we performed additional modeling using conical nanopores with 5 and 7.5 nm in tip radii. As shown in Figure S16, larger conical pores share the same trend as presented for smaller pores in Figure 4c, confirming the generality of our findings.

Figure 4d shows the influence of the base radius on the response time of conical nanopores. At 1 V, the response time increased from 10 to 15 μs when the base



radius increased from 40 nm to 60 nm. When the tip radius is kept constant, the ion concentration inside the conical pore has a weak dependence on the base radius (Figure S17). The increased response time can be then caused by the stronger ionic diffusion from the outside to the inside of the nanopore such that a larger number of ions has to exit the pore to reach a steady state. At −1 V, with the base radius increasing, the electric field strength at the tip becomes stronger, which leads to a larger migration speed of ions (Figure S17d). Also, the decreased degree of ionic enrichment inside the conical pore requires a smaller amount of ions to reach equilibrium (Figure S17b). Under the combined influence of both factors, the response time of conical nanopores decreases with the base radius increasing.

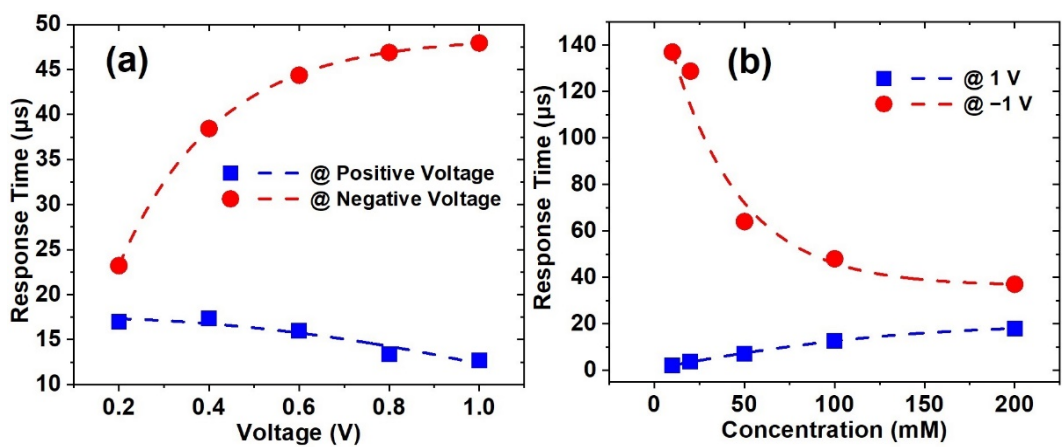

**Figure 5**. Response time of conical nanopores under different conditions, applied voltage (a), and solution concentration (b).

The exploration of conical nanopore response under various external conditions can evaluate the performance of ionic diodes and guide the design and application of high-performance ion diodes. Here, we studied the dynamic regulation of the ion current in the nanopore by voltage and concentration.

The applied voltage is the driving force of ion enrichment and depletion inside



nanopores, whose degree is positively determined by the electric field strength. Figure 5a shows, however, that positive and negative voltages have different effects on the response time of nanopores. For positive voltages, the response time is unchanged and remains at ~15 µs. As shown in Figure S18a, as the voltage becomes more positive, the number of ions that needs to be transported out of the pore increases, while the stronger electric field increases the ionic migration rate (Figure S18c). Therefore, under the combined action of both voltage-induced effects, increased number of ions, and migration rate, the time for the current to reach a steady state is unchanged. At negative voltages, on the other hand, as the voltage increases from 0.2 to 1.0 V, the ionic current takes a longer time to reach its stability, from 23.2 µs to 48 µs. Larger voltages drive more ions into the pore through the tip, increasing ion concentrations (Figure S18b). The enhanced electric field also increases the ion migration rate into the pore (Figure S18d). Both induced changes in ion concentration and ionic migration speed determine the response time of conical nanopores. Due to the more significant enhancement in the number of ions under a larger voltage, the response time of the nanopore increases.

The bulk concentration of the salt determines the thickness of EDLs near charged pore walls,[71] which determines the ion selectivity of the nanopore and affects the regulation of ion migration in the pore by wall charges. Here we consider bulk solution concentrations of 10 to 200 mM. Figure 5b shows that indeed the solution concentration has a significant impact on the response time of ion current. As the concentration increases, the effect of wall charges on ionic transport weakens, and



the degree of ion enrichment and depletion formed inside nanopores decreases. High salt concentrations increase the ion flux at the pore base under positive voltages, and the response time is positively correlated with the increase in solution concentration. Figure S19a shows that with the concentration increase, the number of ions that need to migrate out of the pore tip increases significantly, leading to longer stabilization times. Under negative voltages, on the other hand, high concentrations shorten the response time. This is because the electrostatic shielding effect of the wall is enhanced at higher bulk concentrations, which weakens the cation selectivity of conical pores and overall concentration enhancement (Figure S19b). This allows more ions, especially anions, to enter the conical pore, allowing ion enrichment and ion current to reach stability faster.

We have also performed simulations with the time change in the built-in step function set as 10 µs. The dynamic response of ionic current in conical nanopores was found to share a similar trend under all conditions considered above (Figure S20).

## 4. Conclusions

Due to the asymmetric geometry, conical nanopores with finite surface charges behave as ion current rectifiers. ICR occurs due to the ion enrichment and depletion zones that are created inside conical nanopores at opposite voltage polarities. In this work, with finite element simulations, we explored the time response characteristics of ionic concentrations and ion current in conical nanopores as a function of pore and solution properties. We found that the formation of an "on" – high conductance state



takes a considerably longer amount of time than the formation of an "off" state. At the "on" state, the response time is positively related to the surface charge density, pore length, and voltage, as well as inversely related to the tip and base sizes, and bulk concentration. The response time in the "off" state on the other hand is proportional to the pore length, tip and base sizes, and bulk concentration, as well as inversely proportional to the surface charge density and voltage. The revealed dynamic response of ionic current in conical nanopores can guide the design of high-performance nanofluidic devices based on conical nanopores. In future models, more complex effects such as chemical reactions under nanoconfinement, nanobubbles, and others could open new possibilities for controlling transport at the nanoscale.[29, 58-62]

**Supplementary Information**

See supplementary material for simulation details, additional simulation results of ionic current through nanopores, and ion concentration distributions, electric field strength inside nanopores.

**Author Contributions**

**Zhe Liu:** Simulation, Investigation, Data curation, Writing - original draft, review & editing. **Long Ma:** Data curation. **Hongwen Zhang:** Data curation. **Jiakun Zhuang:** Data curation. **Jia Man:** Investigation. **Zuzanna S. Siwy:** Writing - review & editing. **Yinghua Qiu:** Conceptualization, Methodology, Resources, Writing - original draft, review & editing, Supervision, Funding acquisition.

**Conflict of interest**



The authors have no conflicts to disclose.

## Acknowledgment


This research was supported by the National Natural Science Foundation of China (52105579), the Guangdong Basic and Applied Basic Research Foundation (2023A1515012931), the Natural Science Foundation of Jiangsu Province (BK20200234), the Natural Science Foundation of Shandong Province (ZR2020QE188), and the Qilu Talented Young Scholar Program of Shandong University.